\documentclass{acm_proc_article-sp}
\usepackage{graphicx}
\usepackage{fancyvrb}
\usepackage[toc,page]{appendix}

\begin{document}

\title{Extending the Metasploit Framework to Implement an Evasive Attack Infrastructure}
\subtitle{An exercise for SPICE to test the ability of mainstream antivirus software to prevent
intrusion through drive-by downloads}

\numberofauthors{1} 
\author{
\alignauthor
Aubrey Alston\\
       \email{ada2145@columbia.edu}
\alignauthor
}

\date{6 May 2015}

\maketitle
\begin{abstract}
Given a desired goal of testing the capabilities of mainstream antivirus software
against evasive malicious payloads delivered via drive-by download, the work of this project\footnote{This work is the report corresponding to a project 
done in Columbia University's IDS Lab in Spring of 2015.  All code utilized in this 
project has been archived here: https://github.com/ad-alston/evasive-metasploit} was to extend the functionality of Metasploit--the penetration testing suite of choice--in a 
three-fold manner: (1) to allow it to dynamically generate evasive forms of Metasploit-packaged
malicious binaries, (2) to provide an evasive means of delivering said executables through
a drive-by download-derived attack vector, and (3) coordinate the previous two functionalities
in a manner which can be used to produce reproducible tests within the SPICE framework
\footnote{SPICE is a framework for malware replay and analysis developed and maintained 
privately by the Columbia University Intrusion Detection Systems Laboratory (Columbia 
IDS Lab).}.

As such, the following three major additions to the Metasploit framework were produced:
an evasion applicator module, which, given any Metasploit binary payload, modifies it to
evade static analysis and further modifies the context of its execution through user-
specified templates to achieve evasion of dynamic analysis; a drive-by dropper module
which, given an arbitrary file and a known browser exploit, uploads said file
to the victim's machine using a covert connection; and a multi-dropper module, which is
capable of setting up and managing multiple drive-by dropper instances using the executables
produced by the evasion applicator module.

Following the use of these modules and testing within the SPICE framework
, initial results show an infection rate of over
66 percent over all trials for all tested methods of evasion for the three major
antivirus suites tested (AVG, Avast, and Symantec), and the details derived from
the analysis of results against specific combinations of evasion techniques show
(somewhat alarming) trends regarding the extent and value of the protection provided by
these suites with regard to the chosen attack vector.

\end{abstract}

\section*{Background}

The purpose in creating such an attack infrastructure is simple: to test
the effectiveness of mainstream antivirus software in protecting against
malicious payloads which both attempt to evade analysis and which are placed
without a user's knowledge in a way which is both objective and reproducible.  For 
this reason, the SPICE framework, providing a replay environment for driveby
downloads, is an ideal venue of choice for this realm of exploration.  SPICE
is able to take a URL which may or may not be malicious, instantiate multiple
instances of virtual machines with varying environments, visit that URL, let
what may happen happen, and report changes in the system that occurred as a result
of visiting that URL.  From this, important information may be gleaned: most importantly,
whether or not a machine has been infected.

To create an attack infrastructure, attacks are necessary.  Metasploit is a penetration 
testing suite which is built around an exploit:payload scheme: given a target and an exploit
(and therefore implicitly an attack vector), deliver a payload and gain control of the 
target (generally by acquiring a remote shell [or a platform-agnostic meterpreter shell]).  For this project, Metasploit provides more than sufficient functionality as a 
starting point given its built-in means of delivery detection evasion, also providing
a library of binary executable payloads that are ideal to be objects of an experiment
to test the capabilities of antivirus software.

Aside from the metric infrastructure which SPICE provides and the base
functionality given out of the box by Metasploit, there are areas of 
consideration that are essential to consider: how can a well-documented,
publicly indexed malicious executable like those provided by Metasploit
avoid detection once in place, and how can such an executable be placed
without raising suspicion?  The latter is a question more directly addressed
by the Metasploit framework already,
given that this project specifically assumes the means of placement to 
be drive-by download: given a controlled, exploitable flaw in a browser (Metasploit browser exploit),
simply use that flaw as an attack vector to deliver the executable (modified Metasploit payload).  Network
evasion can be achieved trivially via packet delay, encoding/partial encoding
(e.g. shikata ga nai) of packets, or something as simple as random space injection,
all of which Metasploit provides without touching the baseline.  The prior, interfacing most directly with the focus of this project, is somewhat
more involved.

Reiterating the question whose answer is the key to pushing antivirus software to its limits:
given a possibly known and indexed executable, how can that executable evade notice, detection,
and/or deletion by the host's antivirus?  This question is addressed by evading three primary components of malware analysis: static analysis, dynamic analysis, and reputation-based testing.  Static analysis is analysis of the executable's raw content; dynamic analysis is heuristic analysis of the file's behavior, often achieved through methods like sandboxing; reputation-based testing relies on the number of times a file has been seen, determining whether a file is malicious using a combination of signatures and a database of responses.  The method of addressing this takes inspiration from (and even implements some of the same techniques) the one explored by Eric Nasi: apply encryption-based code armoring by manipulating the compilation process, and wrap decryption and execution in a stub\cite{nasi}.

Nasi's chosen method does have a caveat, however: his encryption scheme requires use of
a somewhat heavy run-time cryptor, which itself may be detected by antivirus software.  Noticing 
this flaw, the given method of evading analysis on the part of the host antivirus software is as follows:
compile code with the encrypted stub already included, opting for a member of a simple family
of light-weight XOR ciphers, and include the decryption and execution stub within a variable
(configurable) context.  In this way, the malicious code is both hidden from static analysis,
and the decryption and execution code can be configured to be different from executable 
to executable as well.  Minimal code is shared between executables, so similarities are
also minimal.  Dynamic analysis evasion is achieved through the context under which
the executable decrypts its malicious content and executes it: for example, the program might
wait x iterations within a for loop to execute in a simplistic case.  Combining these
methods, a well-known, pre-packaged malicious binary can be both modified with regard to both
presentation and behavior in an attempt to pass as a zero-day executable, also providing resiliency against methods of reputation-based testing, since the file's nature as being good or bad can't be necessarily determined given that it's "new."

\section*{The Evasion Applicator Module}

\begin{figure}[h!]
  \centering
    \includegraphics[width=0.5\textwidth]{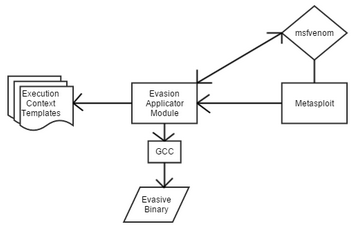}
    \caption{Evasion Applicator Module Architecture}
\end{figure}

(For details on setup, options, and usage, see Appendix A, "Evasion Applicator Module Documentation.")

Given a payload and a set of dynamic evasion techniques and their parameters, the evasion applicator module modifies Metasploit payloads as follows: first, the module delegates to Metasploit's msfvenom utility to output that payload as binary.  Once that output is produced, capture it and read it into memory.  Next, the module modifies that binary to evade static analysis: select from among a family of simple XOR ciphers accepting strings of any length as key, generate a random string as a key--itself having variable length--and encrypt said binary.  Using this encrypted binary and the generated key, build a C program stub which consists of the key, a simple buffer containing the encrypted malicious binary, and the decryption routine corresponding to the chosen XOR cipher (a light-weight, simple nested for loop consisting of 3-5 lines).  Next, read
the list of evasion technique identifiers and their parameters, and interpret their corresponding templates to insert the decryption and execution call within the evasive context(s) [note, for sequentially referenced evasion methods a();b();c(), c's content is executed within b's context, which is itself executed within a's context].  Output the result as a C source file, and then finally delegate to GCC to compile and output a new evasive, executable form of the original binary.

Using the template language defined and accepted by the evasion applicator module, I've
implemented seven\footnote{All of these methods were implemented with a Windows target in mind; only some of them are Windows-specific.} techniques for evading dynamic analysis (though a user of the module
can define as many more or less as he or she likes):

(Note in Appendix A that any evasion technique is recognized by the module as its
identifier plus its parameters (or none, if there are none) in parentheses.)

\subsection*{patience-loop(n)}

This simple method waits n many iterations before executing malicious code.  Antivirus suites
have the task of balancing usability with security, and, as such, have a practical time limit
within which they must make a decision on a newly seen program.  This method attempts
to exceed that time limit.

\subsection*{resource-burn(m,s,r)}

This method allocates m many bytes in s-byte increments, resting r many iterations
between each allocation before executing malicious code.  The reliability of this
method rests on the difficulty of distinguishing between 'normal work' of the average
benign program and, as it is in this case, a distraction from the true malintent 
of the program.

\subsection*{memory-bomb(m, v)}

This method allocates m many bytes, only executing malicious code if the allocation was successful.  The reliability of evading analysis by doing this rests on the assumption that sandboxes are limited in the 
resources they can use.  If the sandbox can't exceed a certain number of bytes 
and/or can't allocate the number requested, the malicious code will not be executed.

\subsection*{who-am-i(e)}

This method verifies that the program was executed as itself (assuming the executable's
name is "e"), only executing the malicious code if this condition is met.  Some sandboxing
systems will invoke a program in other manners which will cause the condition to not be met.

\subsection*{hold-mutex(e, m)}

This method verifies that the program has access to system resources--in this case,
OS-held mutexes.  If the executable (named e) cannot retrieve and hold a named mutex
(of name m), the malicious code will not execute.  The reliability of this method relies
on the assumption that sandboxing which allows a program unhindered access to system resources risks still exposing the host system to attack, hoping that the sandbox limits access to prevent this.

\subsection*{create-file(f)}

This method creates a file at location f and executes malicious code if and only if
the file afterward exists.  The variable file location f allows for this method to
leverage knowledge of host system structure into the attack: if the sandbox allows
creation of a file in a host-specific directory which persists, the host may still
be exposed to an attack.  The reliability of this method lies in the hope that 
access will be limited or emulation will not expose the underlying host's directory structure.

\subsection*{attempt-system-access(p)}

This method executes malicious code if and only if the program is -not- able to
access the process with process number p.  Lower process numbers generally correlate to 
system processes that general applications should not be able to touch, and so if the 
program is successfully able to gain access, the program knows it's being executed with
unexpected permissions and will therefore not execute.

To illustrate the function of this module, consider the following use case:

\verb|msf> use auxiliary/pro/ids/evasion_applicator|\newline
\verb|msf> set EXECUTABLE_NAME example.exe|\newline
\verb|msf> set PAYLOAD windows/meterpreter/reverse_tcp|\newline
\verb|msf> set EVASION_STACK who-am-i(example.exe);|\newline
\verb|       patience-loop(100000)|\newline
\verb|msf> run|

Given this command sequence, the evasion applicator module will then produce an executable with the
following attributes: it will perform the same function as
the windows reverse tcp meterpreter payload without explicitly containing the payload's content, and it will only do so after first
verifying that the program was invoked as 'example.exe' and after iterating 100000
times.  The result is an executable whose intent has already been seen, but whose
presentation and initial behavior present themselves as not having been seen, providing
defense against static, dynamic, and reputation-based analysis.
\vfill\break
\section*{The Drive-by Dropper Module}

\begin{figure}[h!]
  \centering
    \includegraphics[width=0.5\textwidth]{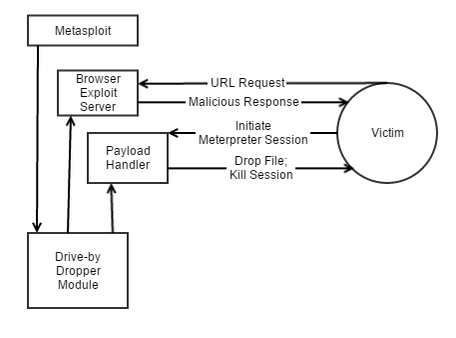}
    \caption{Drive-by Dropper Module Architecture and Attack Scenario}
\end{figure}

(For details on setup, options, and usage, see Appendix B, "Drive-by Dropper Module Documentation.")

The intent of the drive-by dropper module is to create a malicious web server which will drop a configured arbitrary file on the machines of visitors using Metasploit's built-in functionality.  At a high level, the drive-by dropper module achieves this as so: choose a Metasploit browser exploit which allows execution of a payload in memory\footnote{Exploits which do not use browser memory to execute the payload will leave behind artifacts that are easily detected by antivirus software.}, and then choose a compatible meterpreter reverse shell payload.  Setting the corresponding options\footnote{See appendix B.} to these, the module will then coordinate them to achieve the desired effect: delegate to Metasploit to set up\footnote{The module is also written to properly dispose of this (and anything else) it spawns once stopped.} an exploit server using the configured exploit, disabling its default payload handler.  Delegate again to Metasploit to set up a listener to instead handle the incoming meterpreter sessions from successfully executed payloads, itself configured to execute a generated .rc script which uploads the arbitrary file specified onto the victim's machine before killing the covert connection.

Consider the following use case and attack scenario, intended to illustrate the function of the drive-by dropper module:

\verb|msf> use auxiliary/pro/ids/driveby_dropper|\newline
\verb|msf> set EXPLOIT auxiliary/pro/ids/blind_firefox_crmfrequest|\newline
\verb|msf> set DROPDIRS C:/|\newline
\verb|msf> set DROPFILE C:/research/some.exe|\newline
\verb|msf> set DROPNAME some.exe|\newline
\verb|msf> set SHELL windows/meterpreter/reverse_tcp|\newline
\verb|msf> set HOST 127.0.0.1|\newline
\verb|msf> set SERVPORT 8080|\newline
\verb|msf> set LPORT 4000|\newline
\verb|msf> run|

This command will set up a drive-by drop server at\newline http://127.0.0.1:8080/.

Now consider the following scenario:

On a vulnerable machine\footnote{In this case, a Windows user using Firefox version 8 - 14}, somebody visits\newline http://127.0.0.1:8080/.  The browser makes a request to the server, and the
server sends a malicious response which causes an encoded meterpreter session to be initiated from within browser memory to 127.0.0.1:4000.  127.0.0.1:4000 picks up this meterpreter session and
immediately executes a .rc script which navigates to C:/, uploads C:/research/some.exe (on the malicious server) to C:/ (on the victim's computer) as "some.exe", and closes the connection.
The result of visiting this URL: a file being placed without user knowledge or consent.

\section*{The Multi-dropper Module}

\begin{figure}[h!]
  \centering
    \includegraphics[width=0.5\textwidth]{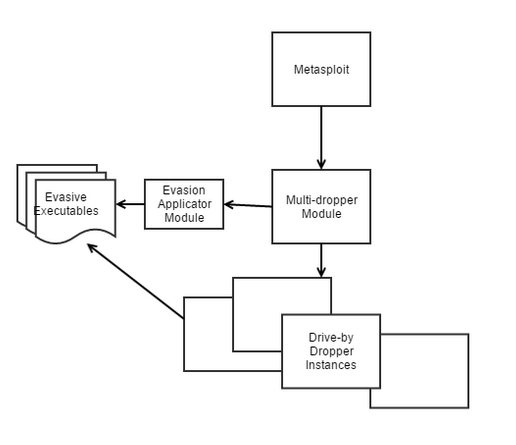}
    \caption{Multi-dropper Module Architecture}
\end{figure}

(For details on setup, options, and usage, see Appendix C, "Multi-Dropper Module Documentation.")

To perform a reproducible test using evasive executables delivered via drive-by download, a coordinating mediator is required: this is the function of the multi-dropper module.  At a high level, the multi-dropper module takes a set of evasion techniques to test, Metasploit payload,
and a configured server location/port range, and (1) generates an executable for each combination of evasion techniques and (2) sets up an instance of the drive-by dropper module for each
for drive-by download delivery.

Consider the following use case, intended to illustrate the functionality of the
multi-dropper module:

\verb|msf> use auxiliary/pro/ids/multi_dropper|\newline
\verb|msf> set EXECUTABLE_PAYLOAD windows/meterpreter/|\newline
\verb|            reverse_tcp|\newline
\verb|msf> set HOST 127.0.0.1|\newline
\verb|msf> set STARTPORT 4000|\newline
\verb|msf> set URIPATH /bad|\newline
\verb|msf> set EVASION_TECHNIQUES a();b();c();d()|\newline
\verb|msf> set DROPNAME some.exe|\newline
\verb|msf> set SERVSTART 1|\newline
\verb|msf> set SERVEND 15|\newline
\verb|msf> set GENERATOR true|\newline
\verb|msf> run\newline|

The multi-dropper module will then create and index one evasive executable
per combination among techniques a(), b(), c(), and d() using the evasion 
applicator module.

\verb|msf> set GENERATOR false|\newline
\verb|msf> run|

Theh multi-dropper will then set up 15 drive-by dropper instances
which will drive-by download the generated executables onto the machine
of vulnerable visitors.

\section*{Using SPICE to Test Antivirus Software}

Our preliminary testing focuses on three mainstream antivirus suites: AVG, Avast,
and Symantec, all of which having considerable market share among consumers of
antivirus software.  The question to be asked is this: just how well do these suites
protect a machine vulnerable to drive-by download?

\subsection*{Method}

To test the ability of AVG, Avast, and Symantec to detect malicious software,
we used the following process:
\begin{enumerate}
  \item Configure SPICE for three virtual machines.  Each is an instance of Windows XP SP2 using Firefox version 11.  One is given AVG; one is given Avast; one is given Symantec.
  \item Use the multi-dropper module to generate evasive executables for all 
  combinations out of a set of evasion techniques applied to the well-known\newline \verb|windows/meterpreter/reverse_tcp| payload.
  \item Use the multi-dropper module to instantiate drive-by dropper instances
  for the generated executables.
  \item Feed the output list of URLs to SPICE, SPICE recording internally
  whether or not each VM is infected by each URL based upon whether the file dropped is
  present after all scanning has taken place.\footnote{We assume a machine to be infected
  if the file still remains because it is possible to induce a user to later try
  to execute that file him or herself.  For example: upload the malicious file
  to the location of and with the name of a file known to exist.  Even if the antivirus
  software has 'blocked' the file from doing anything, if the average end-user assumes
  that the antivirus has somehow made a mistake, he or she may be tempted to disable 
  protection against that file.}
  \item Collect the results from the SPICE database.
 
We tested each antivirus suite against combinations of a set of seven evasion techniques
and configurations, enumerated in figure 4.

\end{enumerate}

\begin{figure}[h!]
    \begin{tabular}{| l | l | p{3cm} |}
    \hline
    \textbf{Technique Configuration} & \textbf{Identifier} \\
    \hline
    patience-loop(10000000000000) & patience* \\ 
    memory-bomb(500000000,01) & memory* \\
    create-file(C:/Program Files/file.tmp) & file* \\
    hold-mutex(metcon.exe,amutex) & mutex* \\
    attempt-system-access(3) & system* \\
    resource-burn(500000,50,10000000000) & burn* \\
    who-am-i(metcon.exe) & who*\\
    \hline
    \end{tabular}
    \caption{Evasion techniques considered}
\end{figure}

\subsection*{Summary of Results\footnote{SPICE is having some problems, so an error state
was induced in some trials.  These results only consider results resulting in an infection
or a block, so marginal proportions may not necessarily agree.}}
\begin{figure}[h!]
    \caption{Detection Rates Among Executables Employing any Combination of Evasion Methods by
    Common Method}
    \begin{tabular}{| l | l | l | l | l | p{3cm} |}
    \hline
     \textbf{Method} & \textbf{AVG} & \textbf{Symantec} & \textbf{Avast} & \textbf{Overall} \\
     \hline
     \textbf{patience*} & 5.38\% & 11.1\% & 65.96\% & 27.58\%\\
     \textbf{memory*} & 18.78\% & 27.64\% & 69.35\% & 38.66\%\\
     \textbf{file*} & 6.63\% & 13.81\% & 47.51\% & 22.65\%\\
     \textbf{mutex*} & 8.94\% & 15.47\% & 50.28\% & 24.95\%\\
     \textbf{system*} & 13.71\% & 20.57\% & 59.43\% & 31.24\%\\
     \textbf{burn*} & 18.23\% & 24.59\% & 67.76\% & 36.93\%\\
     \textbf{who*} & 10.80\% & 19.77\% & 58.19\% & 29.62\%\\
     \textbf{Any} & 12.04\% & 19.95\% & 60.62\% & 30.94\%\\
     \hline
    \end{tabular}
\end{figure}

\begin{figure}[h!]
    \caption{Detection Rates Among Executables Using a Single Evasion Technique}
    \begin{tabular}{| l | l | l | l | l | p{3cm} |}
    \hline
     \textbf{Method} & \textbf{AVG} & \textbf{Symantec} & \textbf{Avast} & \textbf{Overall} \\
     \hline
     \textbf{patience*} & 33.3\% & 28.57\% & 57.14\% & 40.0\%\\
     \textbf{memory*} & 28.57\% & 57.1\% & 100.0\% & 61.9\%\\
     \textbf{file*} & 0.0\% & 0.0\% & 0.0\% & 0.0\%\\
     \textbf{mutex*} & 0.0\% & 0.0\% & 100.0\% & 33.3\%\\
     \textbf{system*} & 0.0\% & 0.0\% & 0.0\% & 0.0\%\\
     \textbf{burn*} & 0.0\% & 33.3\% & 66.7\% & 33.3\%\\
     \textbf{who*} & 0.0\% & 0.0\% & 0.0\% &\% 0.0\%\\
     \hline
    \end{tabular}
\end{figure}

\subsection*{Discussion}

A few observations are immediately apparent from the preliminary results found in
figures 5 and 6.  For example, Avast is doing something AVG and Symantec obviously aren't,
given that the overall detection rate of Avast is three times that of Symantec and nearly six
times that of AVG.  Another observation is that some dynamic evasion techniques are
more effective than others: in figure 5, there is a clear trend among all antivirus software
to have higher-than-normal detection for evasion stacks including the memory-bomb 
and resource-burn methods of evading dynamic analysis, and, likewise, there is a consistent trend for
evasion stacks involving the hold-mutex and create-file methods to have lower-than-average 
detection.  One observation stands out above all others, however: Symantec, Avast, and 
AVG alike all fall significantly short of meeting the 80+\% zero-day detection rates
they are purported to hold, and the executables themselves aren't even true zero-day
malware; they're publicly known, documented pieces of malware that anybody can access 
at any moment.

In addition to this, the content of figure 6 throws uncertainty into the assumption that the
detection rates in figure 5 are anything near a general-case estimate of the capabilities of
these antivirus suites: three attacks consisting of a single evasion method--create-file,
attempt-system-access, and who-am-i--all result in 0\% detection on the part of AVG, Symantec, and Avast.  The non-zero figures in figure 5 can be attributed to the fact that
these undetected methods were used in conjunction with detected ones in at least some
of the combinations tested.  Further, more alarm is raised upon considering the common 
thread held between all three undetected methods: they are all motivated by the assumption
that the sandbox has to be safe.  

Since a sandbox needs to be safe in order to provide any net benefit, by default, it has
to modify the way a program can run and the things it can see, and it has to do so without significantly reducing
usability.  This introduces possible flags as to whether the program is being limited.  In
this case, those flags were whether or not the full filesystem was available, whether or
not the program is given realistic permissions, and whether or not the file is executed
in the precise way in which it is expected: namely, as a program independent of a sandbox. In order for these methods to be detected, the sandboxing/dynamic analysis suite would have to expose elements of risk to the main system.  As an example, exposing the complete structure of the host's directories and files would render create-file useless, but then any program, good or bad, can see this information.  

On a similar key, one has to question what the given detection rates actually mean.  Take any of Avast's higher detection rates: what does Avast's 100\% detection of the memory-bomb technique really mean?  Is it actually detecting the malicious program, or is it blocking a single functionality, assuming that any program which performs it is malicious?  To test this question, I personally wrote a small C program which performed a large allocation of memory (as in the memory-bomb), but without any malicious code or intent included.  I compiled and placed the executable on a virtual machine with the same specifications as the one used in SPICE, and, to little surprise, Avast
flagged the executable as being malicious.

To this extent, then, the initial results of testing popular antivirus software with this
attack infrastructure and SPICE reveal a picture in which nothing is truly certain once the concept of analysis-anticipating malware is brought to the table.  Benign programs are being flagged as malicious for the same reasons as ones which are truly malicious, and truly malicious ones are left undetected because, for an antivirus system to be able to do so, the malware would have to be
allowed to cause actual damage to the host system.  It would appear that the issue of increasing
security against such attacks in the antivirus model calls for one or both of (a) restricting further benign programs or (b) exposing the host system to substantive risk.

\bibliographystyle{abbrv}


\newpage
\onecolumn
\begin{appendices}
\chapter{\textbf{Appendix A, Evasion Applicator Module Documentation}}
\begin{Verbatim}
I. Module Overview

The goal of the evasion applicator module is to take Metasploit binaries and modify
them so they evade static analysis, dynamic sandboxing, and reputation-based
methods of malware detection.  The user invokes the module, specifying
a Metasploit payload and a semi-colon-separated list of template-defined
dynamic evasion execution contexts, and the evasion applicator then
generates a modified binary including (a) the encrypted original binary
and (b) a decryption and execution routine that is carried out within
the context(s) given.

The templates used by the module define a programmatic context in which
malicious code is executed.  In this way, programmatic methods of detecting
sandboxing can be used with any Metasploit payload.

Setup:

The evasion applicator module is a Metasploit module, and as such, must be
added to the framework.

   Steps to add the module to the framework:
       [-.  Install Metasploit.)
       1.  Navigate to your Metasploit installation directory.
       2.  Navigate to the Metasploit modules directory.
       3.  In the "auxiliary" directory, add to or create directories "pro/ids".
       4.  Copy evasion_applicator.rb as well as the "evasion_techniques" folder
           and all of its contents to this directory.

Running instructions:

   1.  Start the Metasploit console.
   2.  Issue command "use auxiliary/pro/ids/evasion_applicator"
   3.  Set options as desired. (Metasploit syntax: set [OPTION] [VALUE])
   4.  Issue command "run".

Options:
	EXECUTABLE_NAME (required) - Specifies the name of the modified binary to
	  be output.

	MSFVENOM_PATH (not required) - Specify the directory in which msfvenom
	  resides (may be necessary for some custom Metasploit installations).

	PAYLOAD (required) - Specify the Metasploit payload to be modified.

	PAYLOAD OPTIONS (not required) - Specify a string of Metasploit options to give to the payload
		(e.g. "LHOST=127.0.0.1 LPORT=4500")

	EVASION_STACK (not required) - Specify a semi-colon separated list of evasion templates
	   to be applied.
	   		Example: technique1();technique2(a);technique3(a,b,c)

	   		Parameters can be passed to the template by specifying them within parentheses
	   		as above.


	LIST_TECHNIQUES (not required) - If this option is set to true, causes the module to
	   output available evasion techniques and their required parameters.

II. Understanding and Adding new Evasion Techniques

The evasion applicator module defines and applies evasion techniques using 
user-defined templates which can be found in the "evasion_techniques" folder.

The module implements and expects a template language adhering to the following
syntax:

[Text block containing comments, usage, etc. The contents of this block
are output by the module whent he LIST_TECHNIQUES option is set.]
%% INCLUDE
[Required header files, one per line]
%% DEFINITIONS
[Definition of functions and constants used in the context section;
standard C syntax expected.]
%%
[C code defining pre-decryption and execution context.]
>> EXECUTE [This is the point where the malicious code is executed.]
[C code defining post decryption and execution context.]
%%

Example scenario:

I know that my target's antivirus software attempts to sandbox all 
files, but a flaw in that sandbox is that 1 + 1 will always evaluate
to 1.  (Unrealistic case.)

---------------------------
example-evasion-technique

Verifies that 1 + 1 == 2 before allowing execution of malicious code.

Usage: example-evasion-technique()
%% INCLUDE
%% DEFINITIONS
#define ONE 1
%%
if(ONE + ONE == 2){
>> EXECUTE
}
----------------------------

A template can access parameters it is passed using @@[param number].

---------------------------
example-evasion-technique

Verifies that a + b == c before allowing execution of malicious code.

Usage: example-evasion-technique(a,b,c)
%% INCLUDE
%% DEFINITIONS
%%
if(@@1 + @@2 == @@3){
>> EXECUTE
}
---------------------------

To add a new template, simply create a file adhering to the proper syntax and
semantics and add it to "[metasploit modules folder]/auxiliary/pro/ids/evasion_techniques".
\end{Verbatim}
\newpage
\chapter{\textbf{Appendix B, Drive-by Dropper Module Documentation}}
\begin{Verbatim}
I. Module Overview

The goal of the drive-by dropper module is to coordinate Metasploit functionality
to configure and create a malicious web-server which will drop an arbitrary
executable in a specified location on the machine of vulnerable visitors.

Setup:

The drive-by dropper module is a Metasploit module, and as such, must be
added to the framework.

   Steps to add the module to the framework:
       [-.  Install Metasploit.)
       1.  Navigate to your Metasploit installation directory.
       2.  Navigate to the Metasploit modules directory.
       3.  In the "auxiliary" directory, add to or create directories "pro/ids".
       4.  Copy driveby_dropper.rb into this directory.

       ** The default option configuration of the drive-by dropper module expects
          an exploit (auxiliary/pro/ids/blind_firefox_crmfrequest) to exist in 
          the framework.  blind_firefox_crmfrequest.rb into this directory as well
          if intending to use the default configuration.

Running instructions:

   1.  Start the Metasploit console.
   2.  Issue command "use auxiliary/pro/ids/driveby_dropper"
   3.  Set options as desired. (Metasploit syntax: set [OPTION] [VALUE])
   4.  Issue command "run".

Options:
	DROPDIRS (not required) - Specifies specific folder(s) in which to drop the
	  arbitrary file.  To configure the module to drop the file in multiple directories,
	  separate directories using a semi-colon.

	DROPFILE (required) - Specify the location of the file to be dropped on the machines
	  of vulnerable visitors.

	DROPNAME (required) - Specify what to name the file on the machine of vulnerable visitors.

	EXPLOIT (required) - Specify the Metasploit browser exploit to use to gain a covert
	  connection.

	HOST (required) - Specify the host/domain the server should use.

	SRVPORT (required) - Specify the port to be used by the server.

    URIPATH (required) - Specify the malicious URI path which will cause the arbitrary file to be
      dropped.

	LPORT (required) - Specify the port to be used in covert connections.

	NETWORK_EVADE - Set to true to configure the dropper to evade network detection.  
	   This causes all meterpreter sessions to be encoded and introduces packet delay
	   and random space injection. **Note: do not use this option with browser exploits
	   which perform probing OS detection.  This can be easily detected.

	SHELL (required) - Specify the meterpreter shell to be used to drop the file. 

	TARGET - Metasploit target code representing the machine of expected vulnerable visitors 
	   (default is 1 - Windows).
\end{Verbatim}
\newpage
\chapter{\textbf{Appendix C, Multi-Dropper Module Documentation}}
\begin{Verbatim}
I. Module Overview

The goal of the multi-dropper module is to coordinate the generation and
distribution of evasive malicious executables on a large, organized scale.
Given a set of evasion techniques, the aim is to create one evasive executable
per combination of evasion techniques and to set up a drive-by dropper 
instance to distribute each as a drive-by download.

When all servers are set up, the module will output two files:
attack_urls, a simple text file containing a list of activated drive-by
dropper instances, and attack_mapping, a line-by-line mapping of each
URL in attack_urls to its corresponding evasion stack.

Setup:

The evasion applicator module is a Metasploit module, and as such, must be
added to the framework.

   Steps to add the module to the framework:
       [-.  Install Metasploit.)
       1.  Navigate to your Metasploit installation directory.
       2.  Navigate to the Metasploit modules directory.
       3.  In the "auxiliary" directory, add to or create directories "pro/ids".
       4.  Copy multi_dropper.rb to this directory.

Running instructions:

   1.  Start the Metasploit console.
   2.  Issue command "use auxiliary/pro/ids/multi_dropper"
   3.  Set options as desired. (Metasploit syntax: set [OPTION] [VALUE])
   4.  Issue command "run".

   * NOTE, evasive executables must be generated before the servers are
     started.  See the 'GENERATOR' option below.

Options:
	DROPDIRS (not required) - Specifies specific folder(s) in which to drop the
	  arbitrary file.  To configure the module to drop the file in multiple directories,
	  separate directories using a semi-colon.

	DROPNAME (required) - Specify what to name the file on the machine of vulnerable visitors
		to drive-by dropper servers.

	EXPLOIT (required) - Specify the Metasploit browser exploit to use to gain a covert
	  connection.

	HOST (required) - Specify the host/domain the servers should use.

	URIPATH (required) - Specify the malicious URI path for the drive-by dropper servers.

	NETWORK_EVADE - Set to true to configure the droppers to evade network detection.  
	   This causes all meterpreter sessions to be encoded and introduces packet delay
	   and random space injection. **Note: do not use this option with browser exploits
	   which perform probing OS detection.  This can be easily detected.

	SHELL (required) - Specify the meterpreter shell to be used to drop the files. 

	TARGET - Metasploit target code representing the machine of expected vulnerable visitors 
	   (default is 1 - Windows).

	EVASION_TECHNIQUES - Set of evasion techniques, the combinations of which to be used
	   to generate evasive executables to be distributed by the drive-by dropper 
	   instances.  (The expected syntax is the same as EVASION_STACK for the evasion
	   applicator module.)

	PAYLOAD_EXECUTABLE - The Metasploit payload to be made evasive and then distributed.

	SERVSTART - (For larger combinations of evasion techniques) - the index of the first
	    combination considered in the range of all combinations.

	SERVEND - (For larger combinations of evasion techniques) - the index of the last
	    combination considered in the range of all combinations.

	STARTPORT - The first port in the range of ports to be used by the drive-by
	    dropper instances.

	GENERATOR - If true, generates the evasive executables and indexes them to be
	    used by the drive-by dropper instances.  If false, sets up the drive-by
	    dropper instances.
\end{Verbatim}
\end{appendices}

\end{document}